\documentclass[aps,twocolumn,pra,showpacs,superscriptaddress]{revtex4}
\usepackage{graphicx}
\usepackage{amsmath}

\newcommand{\be}{\begin{equation}}
\newcommand{\ee}{\end{equation}}
\newcommand{\bea}{\begin{eqnarray}}
\newcommand{\eea}{\end{eqnarray}}

\newcommand{\ket}[1]{\left| #1 \right\rangle}
\newcommand{\bra}[1]{\left\langle #1 \right|}

\newcommand{\mn}[1]{\langle #1 \rangle}

\begin{document}

\title{Simultaneous optical trapping and detection of atoms by microdisk
resonators}

\author{Michael Rosenblit}
\affiliation{Department of Physics and Ilse Katz Center for Meso-
and Nanoscale Science and Technology, Ben Gurion University of the
Negev, P.O.Box 653, Be'er Sheva 84105, Israel}

\author{Yonathan Japha}
\affiliation{Department of Physics and Ilse Katz Center for Meso-
and Nanoscale Science and Technology, Ben Gurion University of the
Negev, P.O.Box 653, Be'er Sheva 84105, Israel}

\author{Peter Horak}
\affiliation{Optoelectronics Research Centre, University of
Southampton, Southampton SO17 1BJ, United Kingdom}

\author{Ron Folman}
\affiliation{Department of Physics and Ilse Katz Center for Meso-
and Nanoscale Science and Technology, Ben Gurion University of the
Negev, P.O.Box 653, Be'er Sheva 84105, Israel}

\date{\today}


\begin{abstract}
We propose a scheme for simultaneously trapping and detecting
single atoms near the surface of a substrate using whispering
gallery modes of a microdisk resonator. For efficient atom-mode
coupling the atom should be placed within approximately 150 nm
from the disk. We show that a combination of red and blue detuned
modes can form an optical trap at such distances while the
back-action of the atom on the field modes can simultaneously be
used for atom detection. We investigate these trapping potentials
including van-der-Waals and Casimir-Polder forces and discuss
corresponding atom detection efficiencies, depending on a variety
of system parameters. Finally, we analyze the feasibility of
non-destructive detection.
\end{abstract}

\maketitle

\section{Introduction}

Optical micro-resonators are currently attracting a lot of
interest in a variety of fields ranging from telecommunication
\cite{slusher} to biological/chemical sensors \cite{vollmer}. In
particular, the advancement of microdisk resonators may lead to
the development of compact and integrable optical-electronic
devices. Recently, significant progress in increasing the finesse
of such resonators has been reported \cite{Kippenberg}, which
makes these devices interesting for future applications in the
emerging field of quantum technology \cite{pelton,qipc,andersson,
atomclock,computer}.

Combining high-Q micro-resonators with miniaturized magnetic traps
for cold, neutral atoms above a substrate, so-called atom chips
\cite{folman}, may lead to integrated devices which allow for a
high degree of control over light-atom interaction. Such systems
may have a significant impact in contexts such as cavity QED
\cite{QED}, single photon sources \cite{pelton}, memory and
purifiers for quantum communication \cite{qipc}, manipulation of
matter waves in interferometric sensors \cite{andersson}, atomic
clocks \cite{atomclock}, and the quantum computer
\cite{computer,qipc}. Such an integrated photonics device for
quantum technology would not only improve technical capabilities
such as enhanced robustness and accuracy while reducing size, cost
and power consumption; it may also give rise to complex new
functionalities such as non-destructive atom-light interaction and
high signal-to-noise detection, high-fidelity qubit transfer and
entanglement for quantum communication, and scalability for, e.g.,
the quantum computer.

Several different realizations of micro-resonators are currently
under investigation in the context of their integration on atom
chips, e.g., Fabry-Perot fiber cavities \cite{horak}, photonic
bandgap structures \cite{lev}, and microdisk resonators
\cite{rosenblit}. The latter possibility is attractive since it
combines the high optical quality of the much studied microsphere
\cite{lefevre,klitzing,kimble} with advanced micro-fabrication and
integration technology.

In this paper we follow up on our recent proposal of using the
whispering gallery modes (WGM) of a toroid microcavity for
single-atom detection \cite{rosenblit}, with a detailed analysis
of the effects of detection on the atomic external degrees of
freedom. In particular, we show that the same optical modes which
are used for atom detection can be exploited to create a trapping
potential for the atom. The parameters can be adjusted to provide
a sufficiently deep potential minimum at an appropriate distance
from the disk surface to hold the atom securely in place during
the detection process.

This work is organized as follows. First, we review the system
under consideration and the principles of optical single-atom
detection in such a device in Sec.\ \ref{sec:mode1}. Next, in
Sec.\ \ref{sec:potential}, we discuss the different optical,
magnetic and surface forces and potentials operating on the atom.
In Sec.\ \ref{sec:trap} we investigate a trapping scheme which
simultaneously allows for optical atom detection. In Sec.\
\ref{sec:detect} we present results on an optimized set of
parameters for simultaneous trapping and detection and discuss the
performance of the detector and the dynamics of the atom during
the detection. Finally, we discuss the experimental feasibility
and conclude in Sec.\ \ref{sec:conclusions}.


\section{System design and operation scheme}
\label{sec:mode1}

\subsection{System design}

The basic system under consideration and its optical properties
have been discussed in detail elsewhere \cite{rosenblit}. Here we
will therefore only give a brief summary.

\begin{figure}
\includegraphics[height=6.5cm]{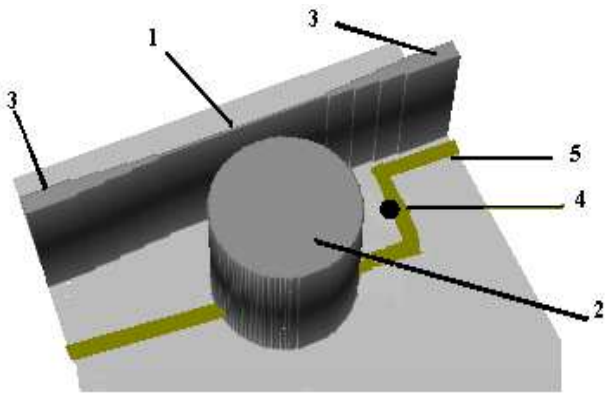}
\caption{The structure under consideration. The evanescent wave
from the slab waveguide (1) is coupled into the disk (2) and back
through a small gap between them. The adiabatic waveguide tapers
(3) serve for coupling light from optical fibers (not shown) into
and out of the waveguide. Cold atoms (4) can be brought to the
disk side via the magnetic field of a $Z$-shaped current-carrying
wire (5) \cite{folman}. However, as explained in this work, such a
magnetic trap is not suitable to hold the atoms during detection,
and so other means of trapping are described for this operation.
\label{fig:structure}}
\end{figure}

We consider an atom chip consisting of a magnetic trap for cold
atoms and an optical resonator for atom trapping and detection as
shown in Fig.\ \ref{fig:structure}.

The optical resonator is a microdisk or a toroid made of a
dielectric material \cite{Kippenberg}, which supports high-finesse
whispering gallery modes near the wavelengths of the D1 and D2
lines of rubidium atoms, i.e., around 795 nm and 780 nm,
respectively. Light is coupled into and out of the cavity through
evanescent-field coupling across an air gap by a tapered linear
waveguide which is mode-matched to achieve best coupling.

Atoms are initially loaded into a magnetic trap \cite{vale} formed
by a $Z$-shaped current carrying wire, as shown in Fig.\
\ref{fig:structure}, with its center located at about 50 - 250 nm
distance from the side wall of the disk. Typically the wire has a
rectangular shape with width and height of 1 $\mu$m, embedded 0.5
$\mu$m below the surface of the chip. Assuming wire currents of
$\sim$100 mA and a homogeneous bias field of 100 G, a magnetic
trap is formed approximately 2 $\mu$m above the surface. However,
as will be discussed later, the magnetic trap is in general too
weak to provide atom confinement during the optical detection
process. The atoms are therefore transferred into an optical trap
formed by the evanescent waves of two microdisk modes of opposite
detuning. The atom-light coupling also changes the optical
properties of the disk modes, which can subsequently be measured
in order to infer the presence of the atom.

The results presented in this work are based on a semi-analytic
coupled mode theory \cite{hammer,love} of the optical properties
of the system and a standard Jaynes-Cummings type model of the
atom-light interaction.

\subsection{Single-atom detection}
\label{sec:detection_intro}

As was described in detail in \cite{rosenblit} our detection
scheme works as follows. Light that is resonant with one microdisk
mode is coupled into the linear waveguide from one side. The
output field at the other end of the waveguide is mixed with a
strong local oscillator field in a balanced detector. The presence
of the atom is then inferred from a change in the intensity
difference between the two arms of the balanced detector.

We have shown previously that, in the limit of far detuning of the
light field from the atomic transition and for low atomic
saturation, the signal-to-noise ratio of this atom detection
scheme is given by
 \be
 S = 4\sqrt{\tau}|A_{in}|\frac{\kappa_T g^2}{\Delta\kappa^2},
 \label{eq:sapprox}
 \ee
where $\tau$ is the measurement time, $A_{in}$ is the amplitude of
the pump light in the linear waveguide normalized such that
$|A_{in}|^2$ is the power in units of photons per second,
$\kappa_T$ is the cavity decay rate due to microdisk-waveguide
coupling, $\kappa$ is the total cavity decay rate including
losses, $\Delta$ is the atom-light detuning, and $g$ is the single
photon Rabi frequency. The corresponding photon number in the
cavity is given by
 \be
 N = 2|A_{in}|^2 \frac{\kappa_T}{\kappa^2}.
 \ee

The optical properties of selected WGMs near resonance with either
the D1 or D2 transition lines of $^{87}$Rb (at 795 nm and 780 nm,
respectively) are summarized in Table \ref{table1} for disk
diameters 15 $\mu$m and 30 $\mu$m. $Q_{1}$ $(Q_{2})$ is the total
quality factor defined by $Q=\omega/(2\kappa)$ \cite{rosenblit}.
As discussed below (Sec.~\ref{sec:trap}) only modes that are blue
detuned with respect to the D2 line and red-detuned with respect
to the D1 line are eventually proposed for detection and trapping.
We assume a root mean square surface roughness of $\sigma=1$ nm
and a surface correlation length of $L_c=5$ nm throughout this
paper \cite{rosenblit}.

\begin{table}
 \begin{ruledtabular}
 \begin{tabular}{ccccccc}
 $D$ ($\mu$m) & $l$ & $\lambda$ (nm) & $Q_1/10^6$ & $Q_2/10^8$ & $g_0$ (MHz) & $\alpha$ (1/$\mu{}$m)\\
  30 & 168  & 774.2  & 3.2  & 1.49 & 100.5 & 7.49 \\
  30 & 167  & 778.73 & 3.0  & 1.47 & 102.6 & 7.7  \\
  30 & 166  & 783.27 & 2.79 & 1.46 & 103.2 & 7.19 \\
  30 & 163  & 797.2  & 2.27 & 1.39 & 105.0 & 7.3  \\
  15 & 82   & 771.3  &1.56 & 0.77 & 202.5 & 7.06 \\
  15 & 79   & 799.2  &1.02 & 0.70 & 209.8 & 6.75 \\
 \end{tabular}
 \end{ruledtabular}
\caption{Optical properties of selected WGMs. $Q_1$ ($Q_2$) is the
quality factor for a waveguide-disk gap size of 0.5$\mu$m
(0.9$\mu$m), $l$ is the longitudinal mode index (the radial index
is $q=1$), $g_0$ is the single-photon Rabi frequency for an atom
at the disk boundary, and $\alpha$ is the decay constant of the
evanescent field. \label{table1}}
\end{table}

Let us now derive an estimate for the light field intensity
required for single-atom detection. Assume we want to detect a Rb
atom using the interaction of the D2 line (780 nm wavelength) with
the $(l=167,q=1)$ mode of a 30 $\mu$m diameter disk, see Table
\ref{table1} for details. The atom is assumed to be located 100 nm
off the surface of the disk for a time $\tau=10$ $\mu$s. In order
to detect the atom with a signal-to-noise ratio of $S=10$, the
required input power is $|A_{in}|^2\approx 0.12\times 10^{14}$
photons per second ($3\mu$W). The corresponding number of photons
in the cavity is $N\approx 2.4\times 10^5$ for a gap size of 0.6
$\mu$m.


\section{Forces and potentials}
\label{sec:potential}

In this section we describe the forces on an atom situated near
the surface of the microdisk: optical forces due to the
interaction with the light fields near the disk, van-der-Waals
forces due to the interaction with the dielectric surface, and
optionally magnetic forces generated by current carrying metal
wires on the surface of the chip. In general, the combined effect
of these interactions is quite complex. In order to simplify the
discussion, we will in the following assume the limit of low
atomic saturation, where the combined potential is a simple sum of
the various contributions,
 \be
 V=V_{\mathrm{light}}+V_{\mathrm{AS}}+V_{\mathrm{mag}}.
 \ee
Here $V_{\mathrm{light}}$ is the optical potential,
$V_{\mathrm{AS}}$ is the atom-surface potential, and
$V_{\mathrm{mag}}$ is the magnetic potential of the wire trap.


\subsection{Atom-light interaction}
\label{sec:light_forces}

In this section we will assume fixed light intensities in the
cavity and only discuss the effects of the light on the atom. The
Hamiltonian of the interaction of a single atom with several
optical modes in the dipole and rotating-wave approximations is
given by
 \bea
    H_{\mathrm{light}}=\frac{i\hbar}{2}\sum_j\sum_{m>n}
    [\Omega_{j,mn}(\mathbf{x})e^{i\omega_j t}|m\rangle\langle n| -
    \nonumber\\
    \Omega_{j,mn}^*(\mathbf{x})e^{-i\omega_j t}|n\rangle\langle m|].
 \eea
Here $|m\rangle$ and $|n\rangle$ are different atomic electronic
levels, and $\Omega_{j,mn}$ is the interaction amplitude between
the disk mode $j$ with frequency $\omega_j$ and the atom
transition $|n\rangle\rightarrow |m\rangle$, such that level
$|m\rangle$ has a higher energy than $|n\rangle$. Note that
different modes may be frequency degenerate (e.g., two
counter-propagating modes) and may couple to the same atomic
transition. In the following we assume coherent states for the
cavity modes and therefore \cite{rosenblit}
 \be
 \Omega_{j,mn}(\mathbf{x}) = 2 g_j(\mathbf{x})\sqrt{N_j}
 \ee
where $N_j$ is the number of cavity photons, $g_j(\mathbf{x})$ is
the single-photon Rabi frequency of mode $j$ for the relevant transition
in an atom at position $\mathbf{x}$ in the evanescent field of the disk mode.
The maximum values of $g_j$ at the disk surface for several
selected modes and the decay constants of their evanescent fields
are given in Table \ref{table1}.

In the simple case of a light field with a single frequency
$\omega$ the Bloch equations for the density matrix of the
two-level atom (with ground state $|0\rangle$ and excited state $|1\rangle$)
fixed at a given point $\mathbf{x}$ have the
steady-state solution
 \bea
 \rho_{11}(\mathbf{x},t) &=&
 \frac{1}{2}\frac{|\Omega(\mathbf{x})|^2}{|\Omega(\mathbf{x})|^2+2\Delta^2+2\Gamma^2}
 \label{eq:rho11}\\
 \rho_{01}(\mathbf{x},t) &=&
 -\frac{\Omega(\mathbf{x})(\Gamma-i\Delta)}{|\Omega(\mathbf{x})|^2+2\Delta^2
 +2\Gamma^2}e^{-i\omega t},\label{eq:rho01}
 \eea
where $2\Gamma$ is the decay rate of the excited state population
by spontaneous emission and $\Delta=\omega-\omega_{01}$ is the
detuning of the mode frequency from the atomic transition
frequency. The light force on the atom is given by
 \be
 \mathbf{F}=-\mathrm{Tr}\{\rho(\mathbf{x})\nabla H_{\mathrm{light}}\}.
 \label{eq:lightforce}
 \ee
If the atomic motion is slow enough such that the change in the
light intensity at the atomic location is small during the decay
time $1/(2\Gamma)$, an adiabatic atomic potential may be defined
which is the spatial integral of the force (\ref{eq:lightforce})
using the steady-state density matrix (\ref{eq:rho11}),
(\ref{eq:rho01}). This yields
 \be
 V_{\mathrm{light}}=\frac{\hbar\Delta}{2}
 \log\left[1+\frac{\Omega^2}{2\Gamma^2+2\Delta^2}\right]\approx
 -\frac{\hbar\Delta}{2}\log[1-2\rho_{11}].
 \label{Vlight_acc}
 \ee
For $\Omega,\Gamma\ll \Delta$ the potential may be approximated by
 \be
 V_{\mathrm{light}}(\mathbf{x})\approx \frac{\hbar
 |\Omega(\mathbf{x})|^2}{4\Delta},
 \label{Vlight_app}
 \ee
which is positive (repulsive) for $\Delta>0$ ("blue detuned"
light) and negative (attractive) for $\Delta<0$ ("red-detuned"
light). Using the values of the parameter estimate of Sec.\
\ref{sec:detection_intro}, we find that the force on an atom 100
nm away from the disk surface is about 70 $\mu$K/nm.

For later use in this paper we now discuss the combined optical potential of
two light fields interacting simultaneously with the atom, where
one field is blue detuned and the other is red detuned. Two
possibilities can be considered: (i) where the two fields couple
the atomic ground state to two different excited states
(three-level situation), or (ii) where both light fields operate
on the same atomic transition (two-level situation).

In the three-level situation, e.g., when the two light fields
operate on the D1 and D2 lines of rubidium, respectively, an exact
analytic solution of the optical Bloch equations for the atomic
density matrix can be found. In the limit of low atomic
saturation, the steady-state 3x3 density matrix may be
approximated by a direct product of the 2x2 matrices for the two
atomic transitions independently. The optical potential is then
found as the sum of two terms given by Eq.\ (\ref{Vlight_app})
corresponding to the two fields.

The situation is slightly more complicated in the two-level case.
Because of interference between the two light fields oscillating
at different frequencies, no steady-state solution exists in this
case. Instead, the atomic density matrix, and hence the optical
force and the dipole potential, oscillate with the difference
frequency $|\Delta_1-\Delta_2|$. However, in the limit of low
atomic saturation it can be shown that the potential oscillates
around a mean value which is the sum of the steady-state
potentials expected from each of the light fields alone. Moreover,
because of the large detuning envisaged here between microdisk
modes and atomic transitions of the order of THz, the oscillation
frequency $|\Delta_1-\Delta_2|$ is much larger than the typical
kinetic energy of the atom. Therefore, the atom will move under an
effective potential which is the time average of the oscillating
potential.

We therefore find that in both configurations, the two-level and
the three-level situations, the optical potential can be
approximated by the sum of two terms of the form
(\ref{Vlight_app}),
 \bea
 V_{\mathrm{light}}(\mathbf{x})
 & \approx &
 V_{\mathrm{light},1}(\mathbf{x})+V_{\mathrm{light},2}(\mathbf{x})
 \nonumber\\
 & \approx &
 \frac{\hbar|\Omega_1(\mathbf{x})|^2}{4\Delta_1}
 +\frac{\hbar|\Omega_2(\mathbf{x})|^2}{4\Delta_2}.
 \eea


\subsection{Atom-surface interaction}

An atom near a dielectric or conducting surface experiences an
effective potential due to the interaction of the atomic dipole
with the dipole moments created in the material. At very short
distances this potential is a van-der-Waals potential due to
static dipole-dipole interaction, while at larger distances of the
order of one wavelength of the atomic transition, the retardation
effect changes the nature of the potential and it is then called
the Casimir-Polder potential \cite{Wu1}. For a ground-state atom
the atom-surface potentials are usually attractive. The asymptotic
form of the atom-surface potential at very short distances from a
non-magnetic and non-dissipative dielectric material with
refractive index $n$ is
 \be
 V_{vdW}(x)=-(\frac{n^2-1}{n^2+1})
  \frac{\mn{d_{\parallel}^2}+2\mn{d_{\perp}^2}}{8\pi\epsilon_0(2x)^3}
 \label{V_vdW}
 \ee
where $d_{\parallel},d_{\perp}$ are the components of the atomic
dipole moments parallel and perpendicular to the surface,
respectively. For an isotropic atom
$\mn{d_{\parallel}^2}+2\mn{d_{\perp}^2} = \frac{4}{3}e^2\mn{r^2}$
where $e$ is the electron charge and $\mn{r^2}$ is the expectation
value of the square of the atomic radius.

The atom-surface potential in the retarded regime, the
Casimir-Polder potential, takes the form
 \be
 V_{CP}(x)=-\frac{\hbar c}{2\pi^2\epsilon_0 (2x)^4}
 \sum_j\frac{c_4^{\parallel}\mn{d_{\parallel}^2}+c_4^{\perp}\mn{d_{\perp}^2}}{E_{ji}},
 \label{V_CP1}
 \ee
where $E_{ji}$ is the energy difference between atomic levels $i$
and $j$, $d_{\parallel},d_{\perp}$ are matrix elements of the
atomic dipole between the two levels at directions parallel and
perpendicular to the surface, and the coefficients
$c_4^{\parallel}, c_4^{\perp}$ are functions of the refractive
index $n$ (see Ref.~\cite{Wu1}) ranging between $0$ for $n=1$ and
$1$ for $n\rightarrow \infty$ (a perfect conductor). We assume an
isotropic atomic dipole moment and use the identity
 \be
 \alpha_0=\sum_j \frac{|\bra{j}d\ket{i}|^2}{E_{ji}}
 \ee
where $\alpha_0$ is the static polarizability of the atom. We then
obtain
 \be
 V_{CP}(x)=-\frac{\alpha_0 \hbar c}{2\pi^2\epsilon_0
 (2x)^4}(2c_4^{\parallel}+c_4^{\perp}).
 \label{V_CP2}
 \ee

The form of the van-der-Waals potential given by Eq.~(\ref{V_vdW})
is valid for very short distances (of the orders of few
nanometers), while the forms of the Casimir-Polder potential,
Eq.~(\ref{V_CP1}) and (\ref{V_CP2}), hold for large distances (of
the order of one micron or more). In the intermediate regime the
exact potential involves cumbersome integrals which may be
performed numerically. For simplicity we make the following
approximation:
 \be
 V_{AS}(x)\approx \int^x dx' \mathrm{max}\{F_{vdW}(x'),F_{CP}(x')\},
 \ee
where $F_{vdW}$ and $F_{CP}$ are the derivatives of the
corresponding potentials in the radial direction.  At short
distances we hence approximate the atom-surface potential by
$V_{vdW}$, while at large distances we use $V_{AS}= V_{CP}$.

\begin{figure}
  \includegraphics[width=6.5cm]{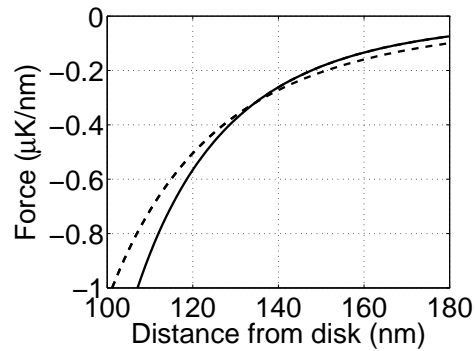}
  \caption{Atom-surface asymptotic forms of the force versus atom distance from the micro
  disk: Casimir-Polder potential (solid line) and van-der-Waals
  potential (dashed line).
  \label{fig:pot2}}
\end{figure}

Figure \ref{fig:pot2} shows the two asymptotic forms of the force
(derivative of the potentials) plotted over the whole range of
distances for a rubidium atom near a silica surface (refractive
index n=1.454). The transition point between these two asymptotic
forms is found at a distance of about 130 nm from the microdisk
surface. The van-der-Waals force at the distance of 50 nm is about
3 $\mu$K/nm and at 100 nm it is about 1 $\mu$K/nm. These values
are therefore much smaller than the repulsive force created by the
blue-detuned detection light as estimated above.

Note that the Casimir-Polder interaction changes with temperature
\cite{Wu2}. However, the temperature dependent correction factor
is mainly important for long atom-surface distances of the order
of a few microns and here we neglect these temperature effects.


\subsection{Magnetic trap potential}

The magnetic field $\mathbf{B}$ created by current carrying wires
on the chip interacts with the magnetic moment $\boldsymbol{\mu}$
of the atom through the Hamiltonian
$H_{\mathrm{mag}}=-\boldsymbol{\mu}\mathbf{B}$. A hyperfine atomic
level with $F>0$ splits in the magnetic field into its Zeeman
sub-levels with magnetic moments following the direction of the
magnetic field while moving adiabatically in the region of the
field. The potential for a specific Zeeman level with quantum
number $m_F$ is
 \be
 V_{m_F}(\mathbf{x})=m_F g_F \mu_B |\mathbf{B}(\mathbf{x})|,
 \ee
where $\mu_B$ is the Bohr magneton and $g_F$ is the Land\'e factor
corresponding to the hyperfine level $F$. Atomic levels with
$m_F>0$ are attracted to the minimum of the field, the $m_F=0$
level feels no potential, and $m_F<0$ levels are repelled from the
minimum of the field. A two-dimensional confinement of levels with
$m_F>0$ is achieved above a straight wire with the help of an
external bias field. Three dimensional confinement is achieved by
the combination of several wires or by a single wire with a $U$ or
$Z$ shape.

A magnetic trap above an atomic chip is usually formed by a
combination of magnetic field components from 3 sources: (a) A
wire in the $y$ direction on the chip surface carrying a current
$I$ generates a magnetic field $\mathbf{B}\approx 2\cdot 10^{-3} I
(z\mathbf{\hat{x}}-x\mathbf{\hat{z}})/(x^2+z^2)$ above the wire
($I$ is in Ampere, $\mathbf{B}$ in Gauss, and $x$, $z$ in meter).
(b) A bias field $-B_0\mathbf{\hat{x}}$ which cancels the magnetic
field generated by the wire at height $z_0=2I/B_0$ from the center
of the wire at $x=0$. (c) An offset field
$B_{\mathrm{offset}}\mathbf{\hat{y}}$, which prevents the magnetic
field at the center of the trap to be zero. The potential near the
center of the trap at $\mathbf{x}_0=(0,0,z_0)$ is then given by
 \be
 \mathbf{B}\approx
 \left((z-z_0)\frac{\partial B_x(\mathbf{x}_0)}{\partial z},
       B_{\mathrm{offset}},
       x\frac{\partial B_z(\mathbf{x}_0)}{\partial x}\right)
 \ee
where $\partial B_x/\partial z=\partial B_z/\partial x \approx
2\cdot 10^{-3}I/z_0^2$. In the range where $(z-z_0)\partial
B_x/\partial z$, $x\partial B_z/\partial x \ll
B_{\mathrm{offset}}$ the potential is approximately harmonic,
 \be
 V_{\mathrm{mag}}(\mathbf{x})\approx m_F g_F
 \mu_B \{B_{\mathrm{offset}}+\frac{2I^2 10^{-3}}{z_0^4
 B_{\mathrm{offset}}} [x^2+(z-z_0)^2]\},
 \label{eq:harmonic_trap}
 \ee
with oscillation frequency
 \be
 \omega_{\mathrm{ho}}=\frac{2\cdot
 10^{-3} I}{z_0^2}\sqrt{\frac{m_F g_F \mu_B}{m
 B_{\mathrm{offset}}}}.
 \label{eq:harmonic_freq}
 \ee
For a $Z$-shaped trap usually $B_{\mathrm{offset}}\approx
B_0\approx 10^{-3} 2I/z_0$. The potential is then harmonic in the
range $|x|,|z-z_0|<z_0$  and the trapping frequency becomes
 \be
 \omega_{\mathrm{ho}}\approx \frac{1}{z_0}\sqrt{\frac{2\cdot 10^{-3}
 I m_F g_F \mu_B}{mz_0}}.
 \ee

Conventional wires can carry a current density of up to $10^7$
A/cm$^2$. For a wire of cross section 1 $\mu\mathrm{m}^2$ this
implies a maximum current of 0.1 A. It follows that for $^{87}$Rb
atoms in the state $|F=2,m_F=2\rangle$ a typical harmonic
oscillator frequency of a magnetic trap can reach up to
$\omega_{\mathrm{ho}}=2\pi\times 35$ kHz at trap heights of
$z_0=3$ $\mu$m above the wire center (2.5 $\mu$m above the top
surface of the wire). The energy splitting between magnetic states
with different $m_F$ ($\Delta m_F=1$) is given by
 \be
 \Delta E\sim\Delta m_F g_F\mu_B |B_{\mathrm{offset}}|.
 \label{eq:splitting}
 \ee
For the strong magnetic trap at a height of 3 $\mu$m the energy
splitting of a $Z$-trap is approximately 50 MHz. The maximal force
that can be applied to the atom by the magnetic field is given by
the magnetic potential gradient, $|F_{\mathrm{mag}}|$ $\approx
2\times 10^{-3}$ $\mu_B$ $I/z_0^2$, which is of the order of 1.5
$\mu$K/nm. This value is much smaller than the force applied by
the evanescent light fields of the disk. This implies that a
magnetic field by itself cannot hold the atom against the
repulsive force of the detection light.

Magnetic fields may still be considered for atom trapping in the
angular direction (along the perimeter of the disk) where no light
forces exist. However, this option is also problematic because of
light-induced Raman transitions into untrapped Zeeman levels.
Raman transitions between different magnetic atomic levels occur
because of stimulated photon absorption and re-emission processes
without significantly populating a higher electronic level of the
atom. In principle, such transitions can be suppressed by
controlling the polarizations of the optical fields. However, in
the case of an evanescent field near the surface of a microdisk
the field polarization cannot be controlled. Raman transitions
will therefore occur with an effective oscillation frequency
 \be
 \Omega_{\mathrm{eff}}\sim \frac{N g(\mathbf{x})^2}{\Delta}.
 \ee
If the value of $\Omega_{\mathrm{eff}}$ between two magnetic
levels is larger than the frequency splitting $\delta$ between the
two levels, significant oscillations will take place between the
levels with flipping times
$t_{\mathrm{flip}}=\pi/2\Omega_{\mathrm{eff}}$. On the other hand
if $\delta>\Omega_{\mathrm{eff}}$, only a fraction
$\Omega_{\mathrm{eff}}^2/(\Omega_{\mathrm{eff}}^2 +\delta^2)$ is
transferred into the other magnetic state. In our system, typical
values of the effective Rabi frequency are of the order of tens of
MHz. This is of the same order as the magnetic splitting of 50 MHz
estimated above, and significant Raman transitions into different
magnetic levels are expected. Note that according to Eq.\
(\ref{eq:splitting}) the magnetic level splitting at the trap
center can be increased, and thus the Raman effect reduced, by
increasing the offset magnetic field, e.g., with the help of
additional current-carrying wires on the chip. However, this comes
at the cost of smaller trap depths and oscillation frequencies,
see Eq.\ (\ref{eq:harmonic_freq}). We therefore conclude that
magnetic trapping in the presence of the evanescent light fields
from the disk would be difficult to achieve in practice, even in
the direction along the perimeter of the disk where no optical
dipole forces exist.


\section{Atom trapping}
\label{sec:trap}

\subsection{Trapping potential}

The discussion in the previous section shows that magnetic and
van-der-Waals forces will in general be too weak to form trapping
potentials for atoms near the microdisk surface in the presence of
the detection light field. In the following we will therefore
investigate atom trapping using \textit{two} light fields: a blue
detuned detection light which also provides a repulsive potential
to keep atoms away from the surface against the atom-surface
force, and an additional red detuned field to provide a long-range
attractive potential which will keep the atoms close enough to the
interaction region with the detection light.

Two objectives must be achieved by the choice of trap parameters.
(i) The trap center must be within 50-200 nm of the disk surface
to allow for atom detection with sufficient signal-to-noise ratio,
e.g., $S>10$. (ii) The depth has to be large enough to keep the
atom trapped for the detection time, taking into account any
heating effects.

\begin{figure}
\includegraphics[width=6.5cm]{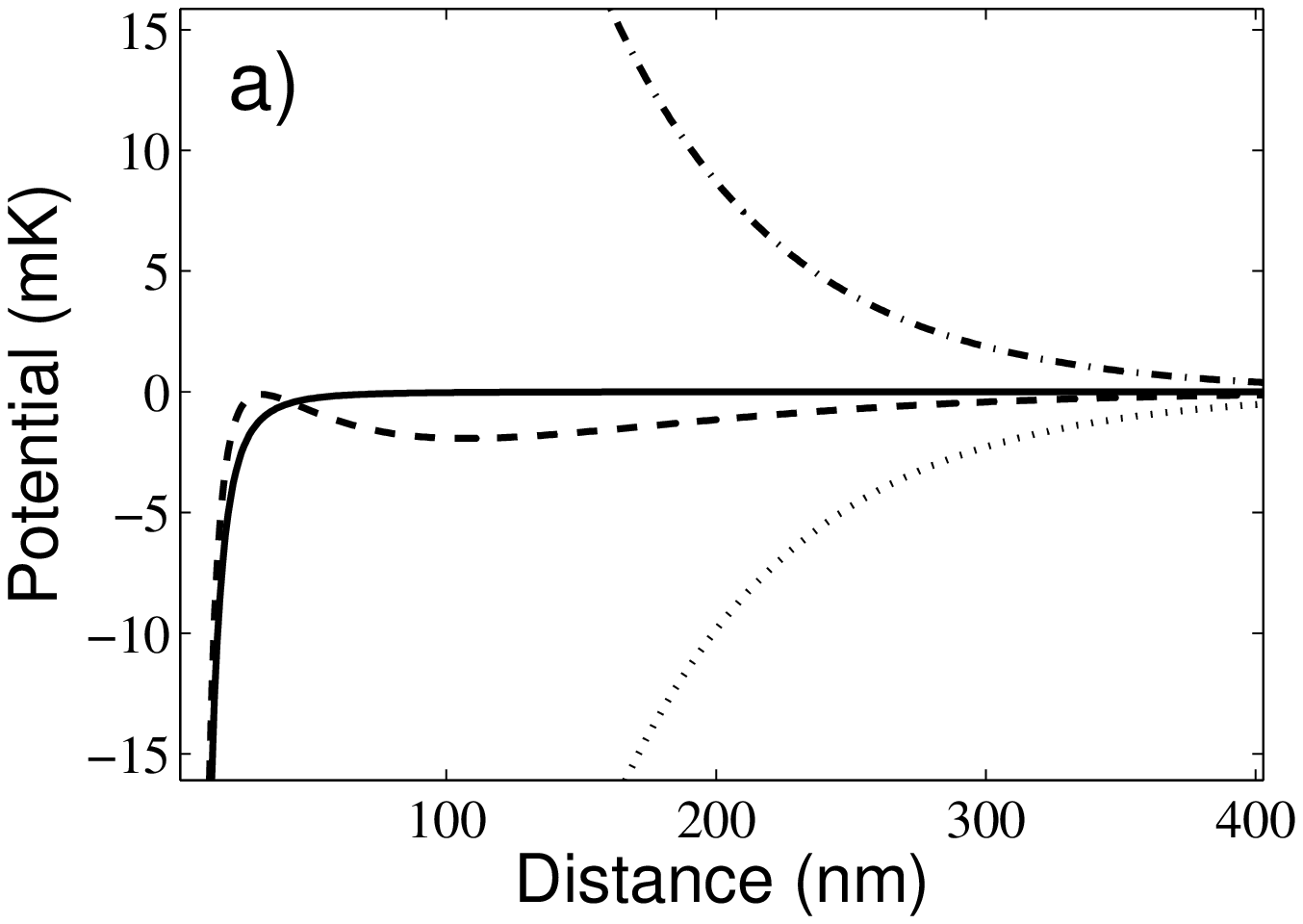}
\includegraphics[width=6.5cm]{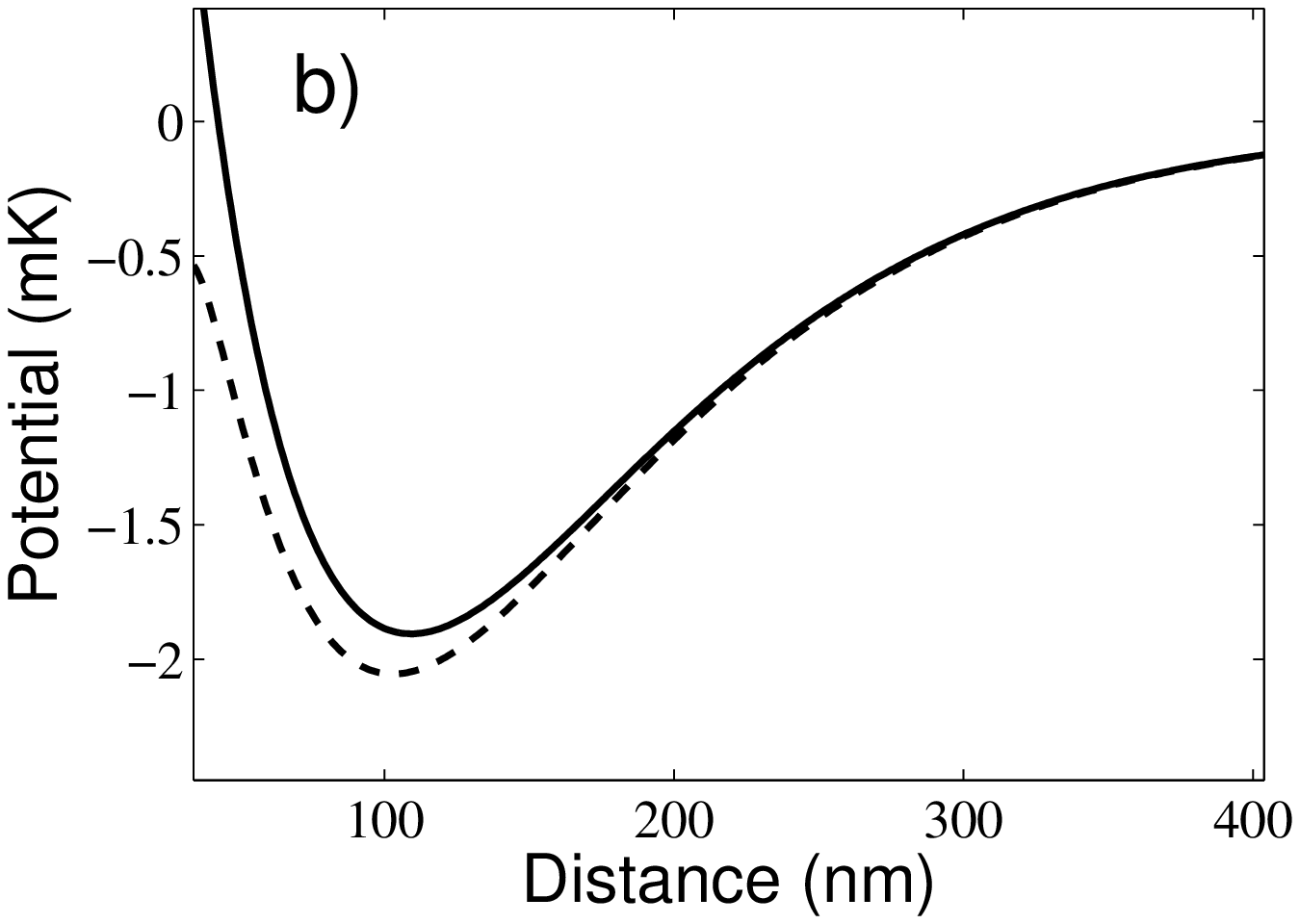}
\caption{The radial potential near the surface of a 30 $\mu$m
disk. (a) Potential formed by blue-detuned light (dash-dotted
line) and red-detuned light (dotted), surface potential (solid),
and sum potential (dashed). (b) The sum of the optical potentials
(solid), and the sum of the optical potentials and the surface
potential (dashed). In this example, the number of blue detuned
cavity photons is $N_b=2.4\times 10^5$, and the number of red
detuned photons is $N_r=3.68\times 10^5$.} \label{fig:potcontrib}
\end{figure}

In Fig.\ \ref{fig:potcontrib} we show sample potentials for a
rubidium atom coupled to the $(l=167,q=1)$ blue detuned mode and
the $(l=163,q=1)$ red detuned mode of a 30 $\mu$m microdisk. The
light intensities create a trapping potential of about 1.9 mK at a
distance of 115 nm from the surface. The surface potential reduces
the potential depth by lowering the potential barrier towards the
disk surface and also slightly shifts the center position.

Before investigating trap optimization numerically in more detail,
we will first discuss a few trap properties in a simple analytic
approximation. For this, we approximate the evanescent fields by
decaying exponentials, i.e., the blue detuned potential is written
as $V_b(r)=V_{b0}e^{-\alpha_b r}$, where $V_{b0}>0$ and $r$ is the
distance from the disk surface, and the red detuned potential is
$V_r(r)=V_{r0}e^{-\alpha_r r}$ with $V_{r0}<0$. The decay
coefficients $\alpha_{r,b}$ are given in Table \ref{table1}. A
minimum of the combined effective potential is formed when the two
corresponding forces cancel. Close to the surface the repulsive
force must dominate, that is, $\alpha_b V_{b0}>\alpha_r |V_{r0}|$.
The minimum is formed at a distance
 \be
 r_{min}=\frac{1}{\alpha_b-\alpha_r}\log\frac{V_{b0}\alpha_b}{|V_{r0}|\alpha_r}.
 \ee
The potential at this point is given by
 \be
 V(r_{min})=-V_{b0}e^{-\alpha r_{min}}\frac{\alpha_b-\alpha_r}{\alpha_r}
 \ee
which is typically a few percent of the single-frequency
potentials. At the minimum, the potential is quadratic with
oscillator frequency
 \be
 \omega_{\mathrm{ho}}=\sqrt{\alpha_r \alpha_b V(r_{min})/m}.
 \ee

Note that the potential is very sensitive to the field
intensities: if $|V_{r0}|$ changes by $\delta V_{r0}$, the
corresponding change in the location of the minimum is
 \be
 \delta r_{min}=\frac{1}{\alpha_b-\alpha_r}\frac{\delta
  V_{r0}}{|V_{r0}|}.
 \label{eq:delta_rmin}
 \ee
Because of the small difference between the two coefficients
$\alpha_b$ and $\alpha_r$ (see Table \ref{table1}), a small change
of the magnitude of the potentials will significantly shift the
minimum of the combined potential. Due to this high sensitivity to
light intensities, a more accurate form of the optical potential
must be examined. For the following calculations, three effects
are taken into account:

(i) The exact form of the potential is given by Hankel functions
\cite{rosenblit} instead of exponentials. Furthermore, atomic
saturation effects render the optical potential nonlinear in the
light intensities, see Eq.\ (\ref{Vlight_acc}).

(ii) The combined potential formed by the two modes is slightly
different than the sum of the two potentials from each mode alone,
cf.\ the discussion in Sec.\ \ref{sec:light_forces}.

(iii) While the van-der-Waals potential is significantly smaller
than the potential of each light field individually, it may still
have an appreciable effect on the sum potential, as the red and
blue detuned fields largely cancel each other.

\begin{figure}
\includegraphics[width=6.5cm]{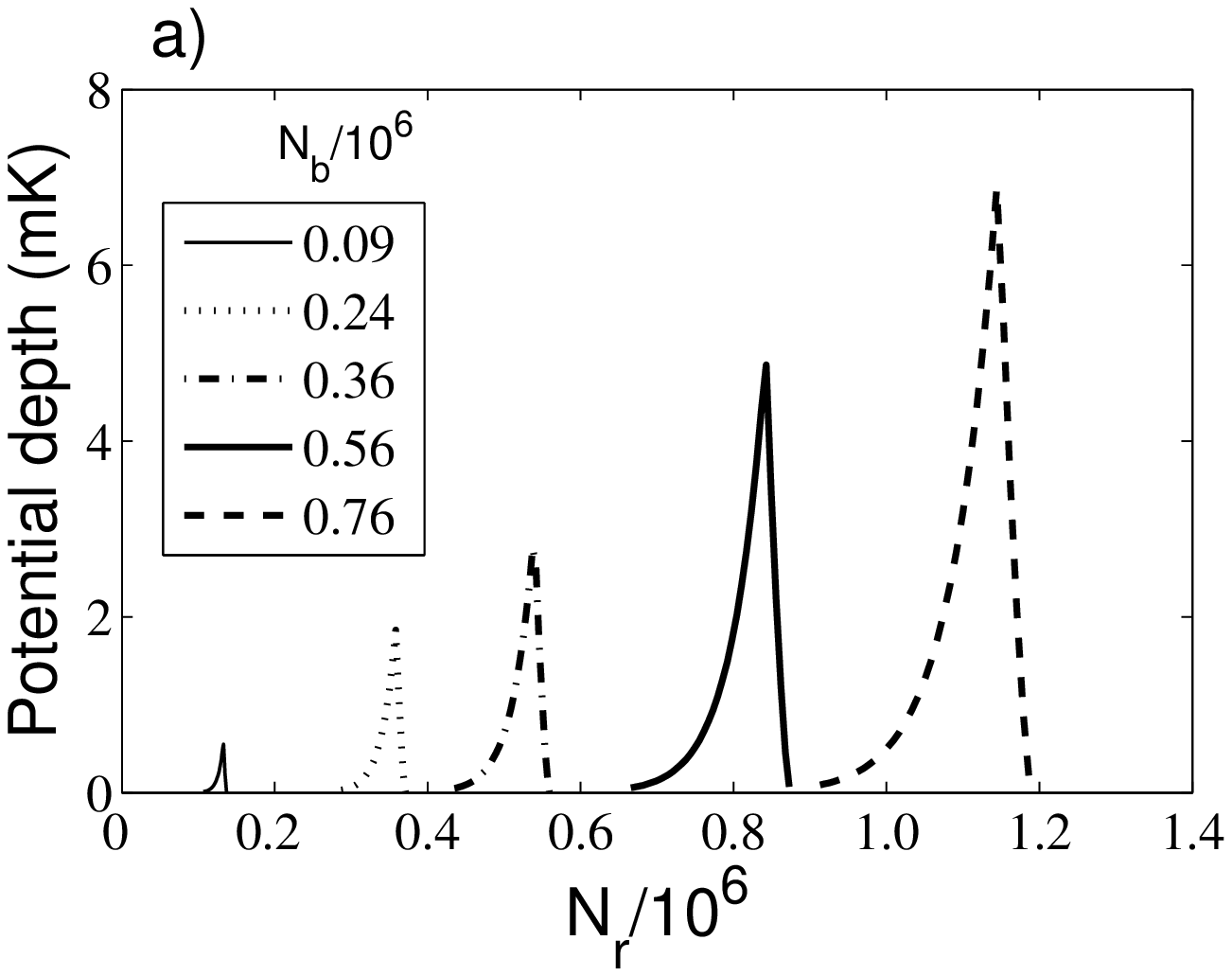}
\includegraphics[width=6.5cm]{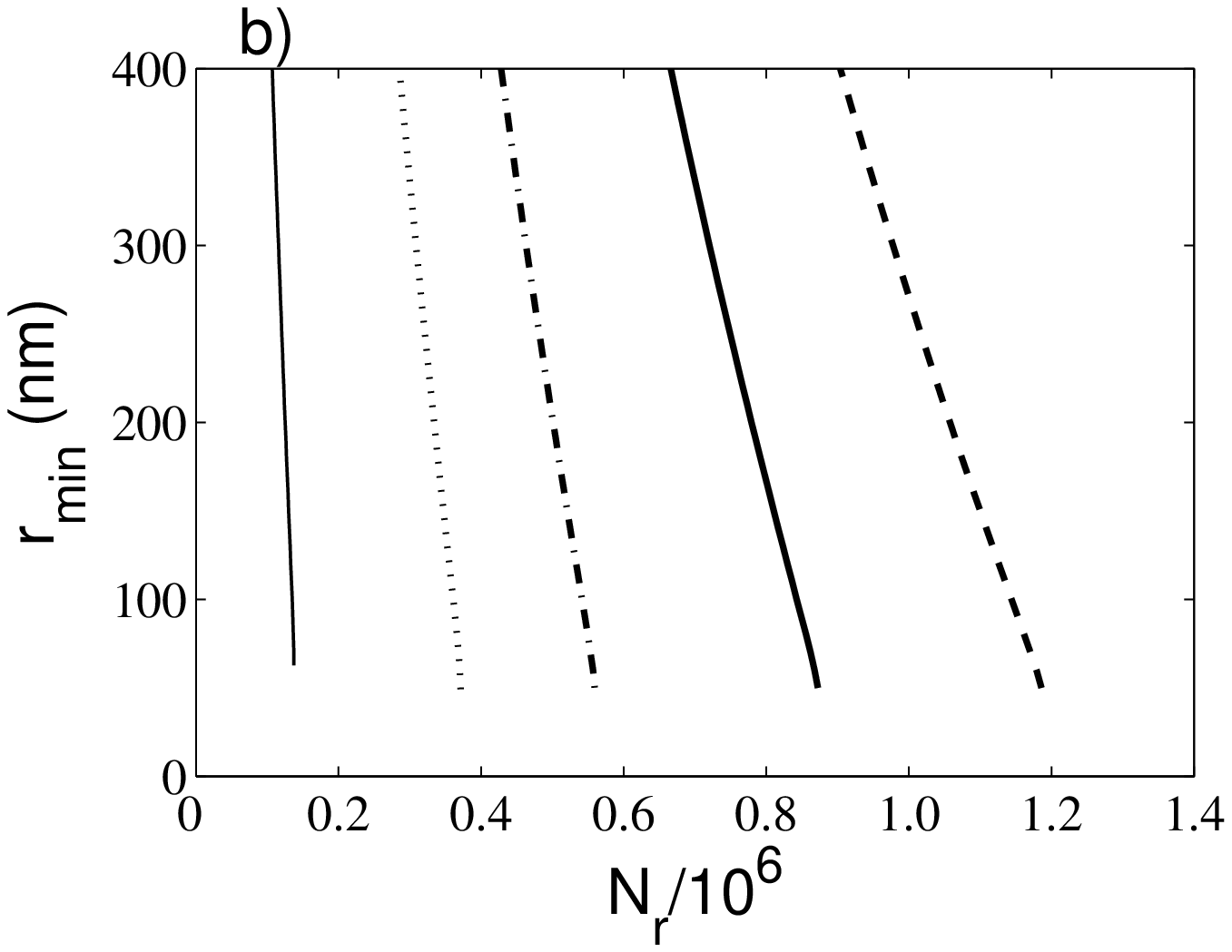}
  \caption{(a) Potential depth and (b) position of the trap
  minimum vs.\ photon number $N_r$ in the red detuned mode for
  different values of $N_b$, the photon number in the blue detuned mode.
  The disk diameter is 30 $\mu$m.
  \label{fig:redvar}}
\end{figure}

The sensitivity of the optical trap to the intensity of the red
detuned light for a fixed intensity of the other mode, as
motivated by Eq.\ (\ref{eq:delta_rmin}), is discussed in Fig.\
\ref{fig:redvar}. The figure shows the potential depth, i.e, the
energy difference between the minimum of the potential and the
maximum towards the disk surface as seen in Fig.\
\ref{fig:potcontrib}, and the position of its minimum versus the
red detuned photon number $N_r$. For too small values of $N_r$, no
trapping potential exists and atoms are either attracted to the
disk surface by the atom-surface interaction or repelled towards
infinity by the blue detuned light. For too large values of $N_r$,
the attractive optical force always dominates over the repulsive
force and all atoms are attracted to the disk surface. Trapping
occurs for a small range of intermediate red-detuned photon
numbers. The maximum potential depth for the chosen parameters is
$\sim 7$ mK and the corresponding center of the trap is at $\sim
120$ nm.

\begin{figure}
\includegraphics[width=6.5cm]{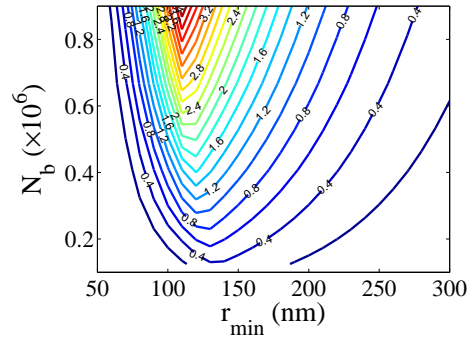}
  \caption{Contour plot of the potential depth (in mK) vs.
  position of the trap minimum $r_{min}$ and photon number $N_b$
  in the blue detuned mode for a disk diameter of 15 $\mu$m.
  \label{fig:redvar15}}
\end{figure}

\begin{figure}
\includegraphics[width=6.5cm]{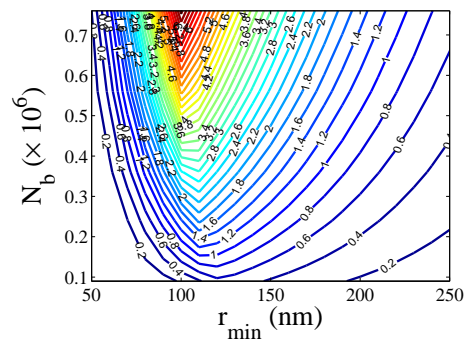}
  \caption{Contour plot of the potential depth (in mK) vs.
  position of the trap minimum $r_{min}$ and photon number $N_b$
  in the blue detuned mode for a disk diameter of 30 $\mu$m.
  \label{fig:redvar30}}
\end{figure}

We summarize the dependence of the potential depth on the
intensity of the detection (blue detuned) light $N_b$ and on the
distance $r_{min}$ for disk sizes 15 $\mu$m and 30 $\mu$m in
Figs.\ \ref{fig:redvar15} and \ref{fig:redvar30}, respectively.
For each pair of values $N_b$ and $r_{min}$ in these figures, the
intensity of the red-detuned light $N_r$ is chosen to yield a trap
center at this position $r_{min}$.

Note that by increasing the intensity of \textit{both} light
fields arbitrarily deep optical potentials can be created in
principle. However, this will also change the detection efficiency
and increase the back action on the atom itself, as will be
discussed later.

\subsection{3D trapping}

So far, we have only discussed trapping in the radial direction
from the disk surface. However, we would like to ensure full
three-dimensional trapping. First, this would enable us to use an
atom again after detection and furthermore, tight 3D confinement
would reduce atomic heating as discussed in Section
\ref{sec:heat}.

The optical potential that is formed by the evanescent field of
the WGMs in the disk also depends on the $z$ coordinate along the
height of the disk. The field intensity along this axis can be
approximated by $V(z)\approx V(H/2)\cos^2[\pi(1-2z/H)/2]$, where
$H$ is the height of the disk \cite{jp}. Thus, the intensity
vanishes at the edges $z=0$ and $z=H$ of the disk. In the center,
at $z=H/2$, a potential minimum is formed with harmonic oscillator
frequency
 \be
 \omega_{z,\mathrm{ho}}=\frac{\pi}{H}\sqrt{|V_{\mathrm{max}}|/m}.
 \label{eq:omegaz}
 \ee
This frequency is typically a few kHz, which is an order of
magnitude smaller than the radial trapping frequency.

Confinement in the angular direction may be achieved if we create
a red-detuned standing wave by inputting red-detuned light from
the two sides of the linear waveguide equally to create a
superposition of clockwise and counterclockwise propagation around
the disk. In this case the red-detuned and blue-detuned light
potential will be given by
 \bea
 V_{\mathrm{red}}(x,y,z) & = &
    V_{\mathrm{red}}(0,0,H/2)e^{-\alpha_r x}\\
 & & \times \cos^2(\pi(z-H/2)/H) \cos^2(l_ry/R)  \nonumber\\
 V_{\mathrm{blue}}(x,y,z) & = &
    V_{\mathrm{blue}}(0,0,H/2)e^{-\alpha_b x} \\
 & & \times \cos^2(\pi (z-H/2)/H)  \nonumber
 \eea
where $l_r$ is the winding number of the red-detuned WGM and
$R=D/2$ is the microdisk radius. The harmonic oscillator frequency
of the trapping potential in the $y$ direction is then
 \be
 \omega_{y,\mathrm{ho}}=\frac{l_r}{R}\sqrt{|V_{\mathrm{red}}^{\mathrm{max}}|/m}.
 \ee
where $V_{\mathrm red}^{\mathrm max}$ is the red-detuned potential
at the center of the trap.

Finally, in order to ensure trapping we need to ensure that the
tunneling probability to the disk is negligible,  For typical
values of barrier height of 1-2 mK and barrier width of 30-60 nm
we expect that no significant tunneling will occur in a time of
less than few seconds. In order to keep these values of barrier
height and width the trapping distance of the atom from the
surface  must be larger then $\sim 80$ nm.


\subsection{Heating of a trapped atom by spontaneous emission}
\label{sec:heat}

Interaction of the atom with the trapping and detection light
fields will in general also induce heating of the atom. Two
heating mechanisms can be distinguished: fluctuations of the
dipole forces due to, for example, mechanical vibrations or laser
instability, and recoil heating after spontaneous emission events
\cite{vern}. Here we will focus on the latter heating mechanism.

In a shallow potential well, each spontaneously emitted photon
will on average add one photon recoil energy $E_r= \hbar^2
k^2/(2m)$ to the kinetic energy of the photon, where $k$ is the
wave-vector of the photon. Thus, if we define
 \be
 M = 2\Gamma\tau\rho_{11}
 \label{eq:M}
 \ee
as the number of spontaneous emission events during the atom
detection time $\tau$, the total heating is given by $ME_r$. We
therefore require potential depths $V>ME_r$ in order to hold an
initially ultra-cold atom in the trapping potential during the
interaction with the light fields.

In a steep potential well, on the other hand, atom trapping is
aided by the Lamb-Dicke effect: an atom in the oscillatory ground
state is much more likely to return to the same state after a
spontaneous emission than to be excited to a higher oscillation
state. In the following we will derive an expression for the
probability of the atom to remain in the ground state of motion
during the detection time.

Let us assume that the potential may be approximated as a harmonic
potential $V=\frac{1}{2}m (\omega_x^2 x^2+\omega_y^2
y^2+\omega_z^2 z^2)$ with ground state sizes $x_0$, $y_0$, $z_0$
in the three directions, respectively. The probability that an
atom in the harmonic ground state $|\psi_0\rangle$ returns into
the same state after the emission of a photon with wave vector
$\mathbf{k}$ is given by
 \be
 P_0(\mathbf{k})=|\bra{\psi_0}e^{i(k_x x+k_y y+k_z z)}\ket{\psi_0}|^2.
 \ee
Averaging over all possible directions of $\mathbf{k}$ yields
 \be
 P_0=\frac{\sqrt{\pi}}{2}\frac{\mathrm{erf}(kr_0)}{kr_0},
 \ee
where
 \be
 r_0=\sqrt{x_0^2+y_0^2+z_0^2}=\sqrt{\frac{\hbar}{2m}\left(\frac{1}{\omega_x}
 +\frac{1}{\omega_y}+\frac{1}{\omega_z}\right)}.
 \ee
For $kr_0\ll 1$ this may be approximated by
 \be
 P_0\approx 1-\frac{k^2 r_0^2}{3}=1-\frac{\hbar
 k^2}{6m}\left(\frac{1}{\omega_x}
 +\frac{1}{\omega_y}+\frac{1}{\omega_z}\right).
 \ee

After $M$ spontaneous emissions the probability of staying in the
ground state is $P_0^M$ and thus the probability of scattering
into excited states of motion is given by
 \be
 P_{\mathrm{other}}(\tau)=1-P_0^M \approx 2\Gamma\rho_{11}\tau
 \frac{\hbar k^2}{6m}\left(\frac{1}{\omega_x}
 +\frac{1}{\omega_y}+\frac{1}{\omega_z}\right).
 \label{pother}
 \ee
If this probability is much smaller than unity, the atom will stay
in the ground state during the detection process with high
probability.


\subsection{Trapping stability}

\label{sec:back}

Bi-chromatic atom trapping has already been used in several
applications, beginning from the work of Ovchinnikov \textit{et
al.} \cite{ovchinnikov}, where two colors with different
evanescent decay lengths have been proposed in order to trap atoms
at a distance $~\lambda$ from a prism surface. The two color
scheme has been considered also for a dielectric microsphere, a
free-standing channel waveguide and an integrated optical
waveguide \cite{vern,mabuchi,barnett,burke,lekien}. The
back-action of an atom on the bi-chromatic light field in the
strong coupling regime was considered for atom trapping and
cooling in Ref.\ \cite{domrit}.

The main disadvantage of bi-chromatic trapping, as was pointed out
by Burke \textit{et al.} \cite{burke}, is that trapping is
achieved by a fine balance between two optical fields, such that
even a small intensity fluctuation may drastically change the
trapping conditions or even destroy the trap. Such spatial
intensity fluctuations usually exist in real waveguides due to
imperfections that generate backscatterred waves which interfere
with the main light wave. It was shown \cite{burke} that even a
backscattered wave, whose intensity is only $0.001$ of the
propagating mode intensity, may decrease the potential depth by a
half and consequently destroy the trap.

Here we analyze the intensity spatial fluctuations due to
backscattering in the microdisk structure. We show that
backscattering may be decreased to a level where such fluctuations
will not severely change the trapping conditions. This is
achievable by increasing the coupling rate between the linear
waveguide and the microdisk. This increase may affect the
detection signal-to-noise ratio and consequently the integration
time will need to be increased.

As was described in \cite{rosenblit} the linear coupling between
the two counter-propagating modes is described by a complex
coefficient $\epsilon$, which is related to the intrinsic loss
rate of the disk $\kappa_{int}$ due to imperfections. The total
loss rate of the disk $\kappa$ is a sum of this intrinsic loss and
the loss $\kappa_T$ due to waveguide - disk coupling
\cite{rosenblit}, such that $\kappa_{int}=\kappa-\kappa_T$. In
this model the mode amplitudes are given by complex numbers
$\alpha_+$ and $\alpha_-$, respectively, such that
$|\alpha_{\pm}|^2=N_{\pm}$ are the numbers of photons in the two
modes. If the disk is pumped with rate $\eta$ then the steady
state values of $\alpha_{\pm}$ are given by \bea \alpha_-&
=&(\epsilon/ \kappa)\alpha_+ \\ \alpha_+& =& \frac{ \eta}{\kappa (
1+\epsilon^2/\kappa^2)} \eea so that the backscattered to
propagating ratio is given by
 \be (I_- / I_+)^{1/2} =
 \frac{\epsilon}{\kappa-\kappa_T}\frac{\kappa-\kappa_T}{\kappa}
 =\frac{\epsilon}{\kappa_{int}}\left(1-\frac{\kappa_T}{\kappa}\right)
 \ee

With high-quality microdisk resonators,
$\epsilon/\kappa_{int}\approx 1$ can be achieved
\cite{Kippenberg}. Using this value we find that if we wish to
keep the trap depth fluctuations smaller than $\pm 2\%$ (meaning
$(I_-/I_+)^{1/2}<1\%$ \cite{equation}), we require
$\kappa_T/\kappa>0.99$. This value may be achieved by decreasing
the gap between waveguide and microdisk to $\sim 0.46\mu$m in both
disks of diameters $D=30\mu$m and $D=15\mu$m, with corresponding
$Q$ values of $Q\approx 1.7\times 10^6$ and $Q=0.9\times 10^6$,
respectively. In this case, $I_-/I_+=10^{-4}$, i.e., an order of
magnitude better than that considered by Burke et al., which leads
to acceptable trapping instabilities. Thus we believe the concerns
of Burke et al. described earlier, regarding the destruction of
the trap due to back scattered light, are not warranted for
state-of-the-art micro-disks.


\section{Detection of trapped atoms}
\label{sec:detect}

Having discussed the principles of atom detection (Sec.\
\ref{sec:mode1}) and trapping (Secs.\ \ref{sec:potential} and
\ref{sec:trap}), we will now combine all the components and find
an optimized set of parameters for efficient simultaneous atom
detection and trapping.

We consider the detection and trapping of an atom in a three-level
configuration with two light fields. Two disk sizes are
investigated: a 30 $\mu$m disk using the ($l=167,q=1$) mode as the
blue detuned detection light and the ($l=163,q=1$) mode as the red
detuned attractive trapping light, and a 15 $\mu$m disk with the
($l=82,q=1$) mode as the blue-detuned detection light and the
($l=79,q=1$) mode as the red-detuned attractive trapping light.
The optical properties of these modes are given in Table
\ref{table1}. For both configurations we assume surface quality
parameters of $\sigma=1$ nm and $L_c=5$ nm, and a gap size of
$\sim$ 0.5 $\mu$m.

\begin{figure}
\includegraphics[width=6.5cm]{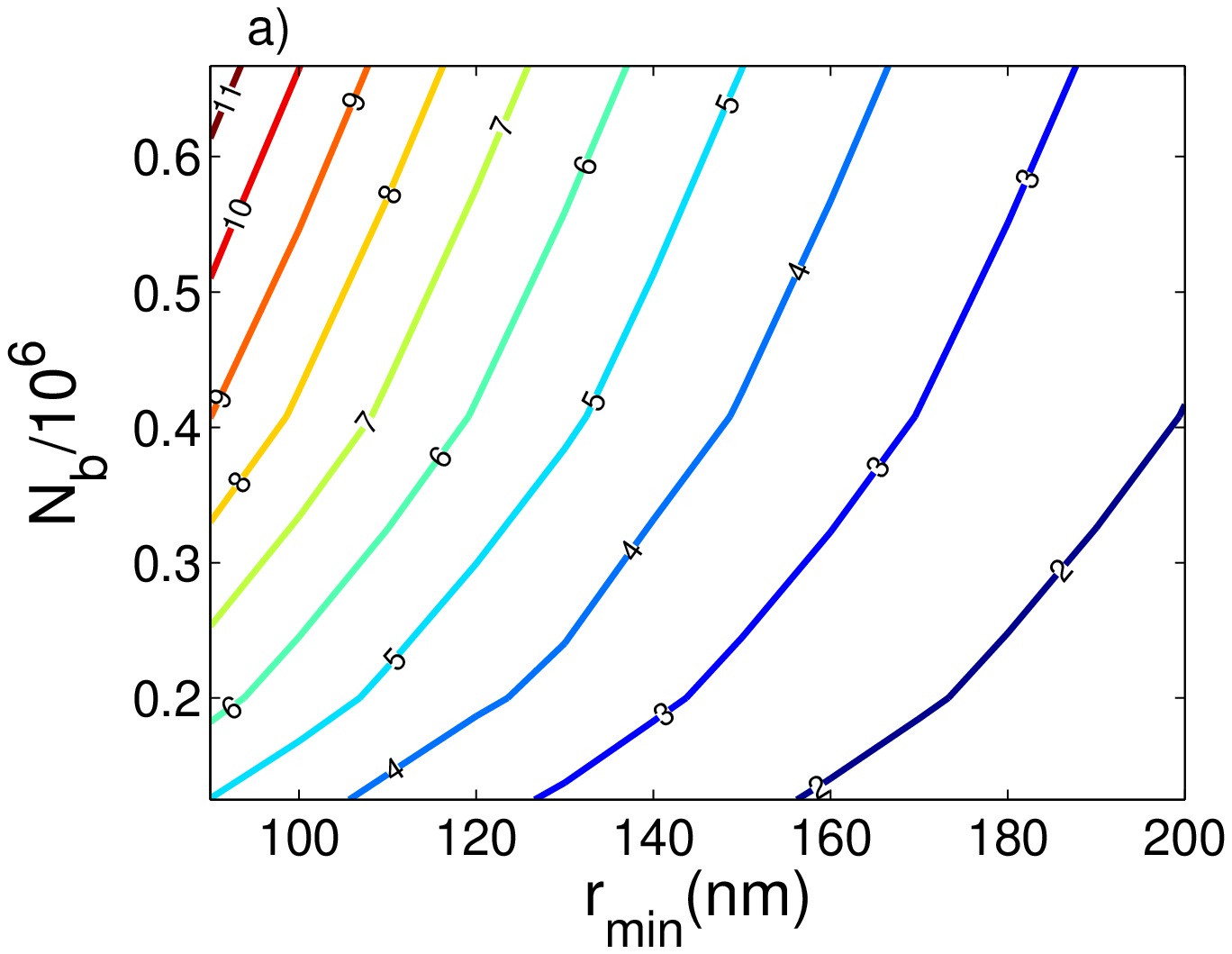}
\includegraphics[width=6.5cm]{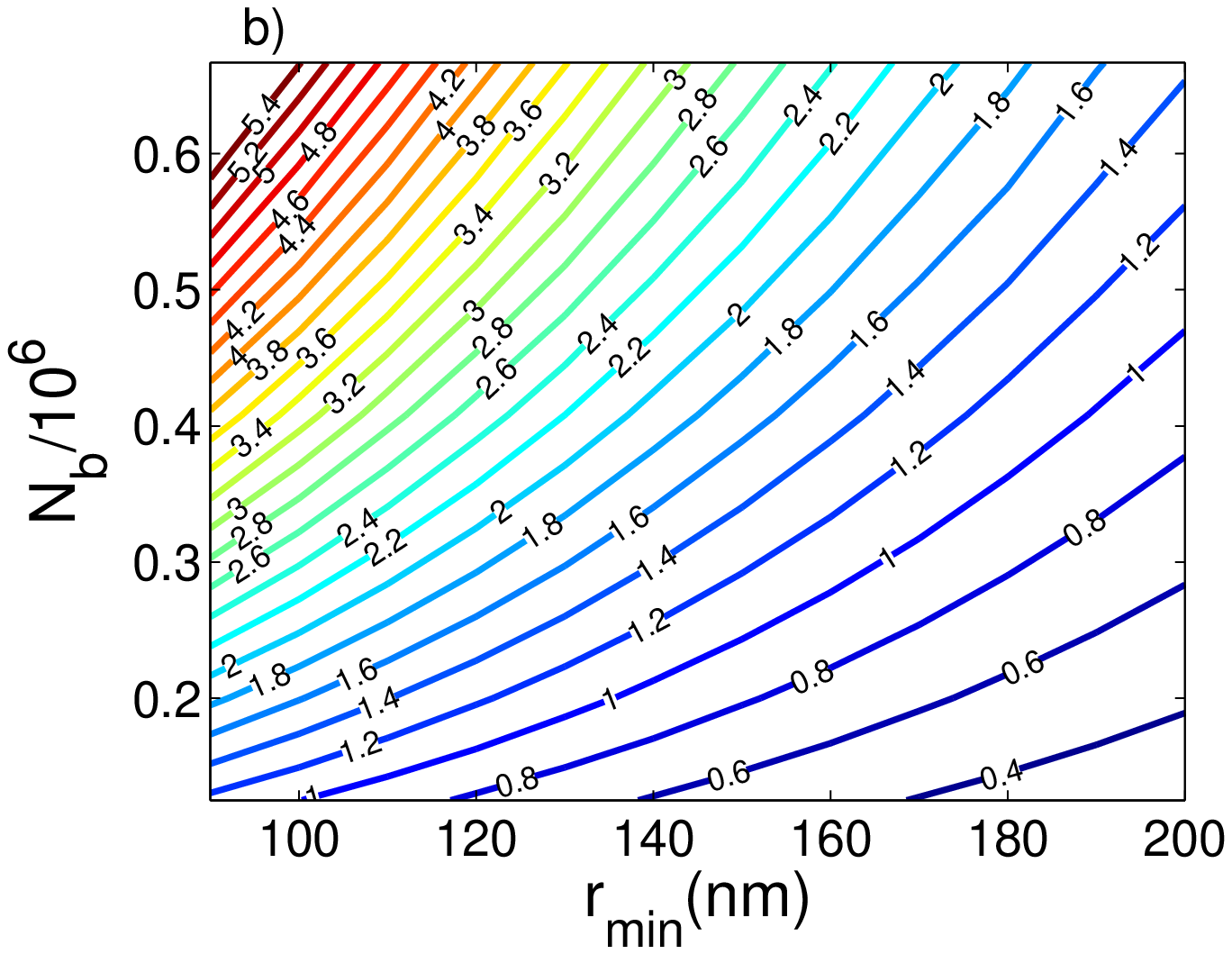}
  \caption{(a) Signal to noise ratio and (b) total (red and blue) scattered photon number
  vs. distance from the disk and blue detuned photon number for a disk
  diameter of 15 $\mu$m. The integration time is 125 $\mu$s.
  \label{fig:SM15}}
\end{figure}

\begin{figure}
\includegraphics[width=6.5cm]{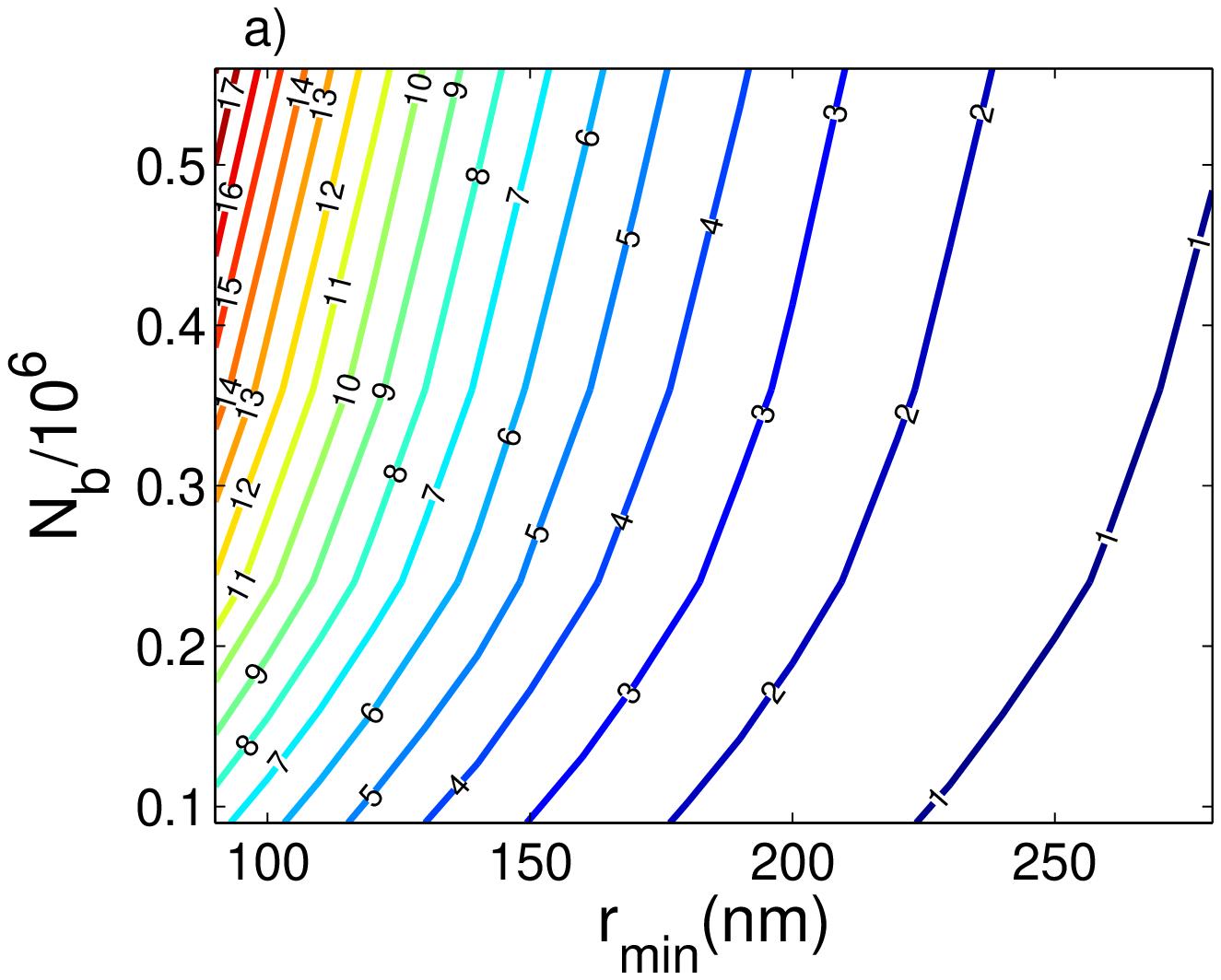}
\includegraphics[width=6.5cm]{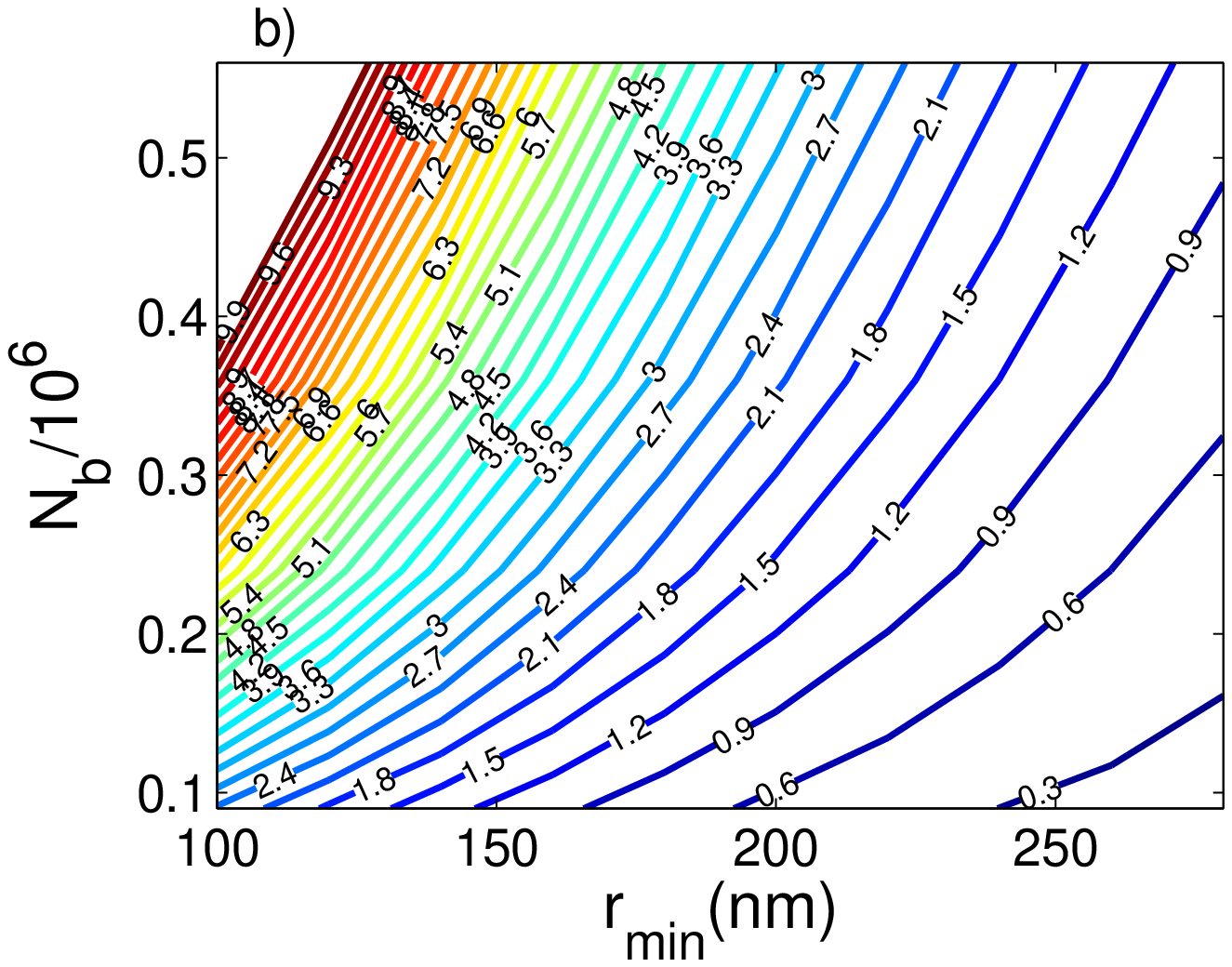}
  \caption{(a) Signal to noise ratio and (b) total scattered photon numbers
  vs. distance from the disk and blue detuned photon number for a disk diameter of 30 $\mu$m.
  The integration time is 75 $\mu$s.
  \label{fig:SM30_vsd}}
\end{figure}

We start our parameter optimization procedure by investigating the
single-atom detection efficiency as a function of light
intensities. To this end we show in in Figs.\ \ref{fig:SM15}(a)
and \ref{fig:SM30_vsd}(a), contour plots of the signal-to-noise
ratio $S$, versus blue detuned photon number in the disk and
versus the position $r_{min}$ of the trap minimum from the disk
surface (this is shown for disk diameters 15 and 30 $\mu$m with
observation times $\tau=125$ and 75 $\mu$s respectively). As
expected, $S$ increases with increasing light intensity, because
of the better photon statistics, and with decreasing atom-disk
distance, because of increasing coupling constant $g(\mathbf{x})$.

Next, we calculate the intensity of the red-detuned light required
to achieve a trapping potential minimum at given values of
$r_{min}$. From this we obtain the total number of photons $M$,
defined in eq.\ (\ref{eq:M}), spontaneously scattered out of the
two light fields by the atom during the detection process. The
results show that, in general, larger values of $S$ imply larger
numbers of scattered photons $M$ as depicted in Figs.\
\ref{fig:SM15}(b) and \ref{fig:SM30_vsd}(b). However, we note that
the parameter dependence of $S$ and $M$ is different, so that
optimization is possible.

\begin{figure}
\includegraphics[width=6.5cm]{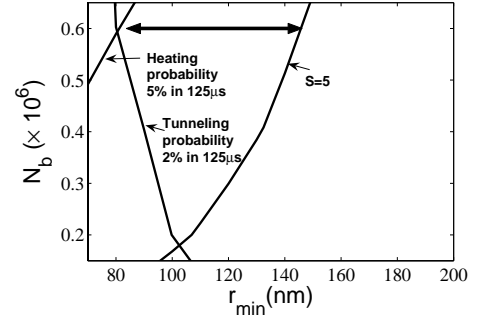}
\caption{Performance characteristics of a disk with diameter
$D=15$ $\mu$m and for a $125\mu$s integration time, at different
blue-detuned photon numbers and atom distances. The area confined
by the lines represents $S>5$, heating probability $<5\%$ and
tunneling probability of $<2\%$ contains the parameter region
where stable atom trapping and detection can be achieved with high
confidence. An analysis of the parameter tolerance shows that for
these optimized parameters a change of $\pm$2\% in either the blue
or red-detuned light intensity will still lead to another point
within the triangle. The arrow represents the change in trap
position if the intensity of the red- or blue-detuned light is
varied by $\pm 2\%$. \label{fig:merit15}}
\end{figure}

In Fig.\ \ref{fig:merit15} we combine for a $D=15$ $\mu$m disk the
contour lines for an $S=5$ detection, a heating probability of
5\%, Eq.~(\ref{pother}), and a tunneling probability of 2\%,
calculated using the standard expression based on the WKB
approximation. The optimal point for atom detection within the
parameter region of Fig.\ \ref{fig:merit15} is found in the middle
of the triangle for $N_b=6 \times 10^5$ and a trap at distance of
$r=115$ nm from the disk. In order to achieve atomic trapping at
this point we need $2.5\times 10^5$ red-detuned photons in the
disk. At this point we obtain $S=8$. The potential depth is 2.6 mK
and the harmonic oscillator frequency is $(\omega_{\mathrm{x}},
\omega_{\mathrm{y}},\omega_{\mathrm{z}})= 2\pi\times (1.5,4,0.14)$
MHz. The ground state energy of a single atom in this trap is
approximately 160 $\mu$K and the ground state dimensions are
$(\Delta x,\Delta y,\Delta z)=(15,6.5,40)$ nm. The heating
probability is $4\%$. An analysis of the parameter tolerance shows
that for these optimized parameters a change of $\pm$2\% in either
the blue or red-detuned light intensity will still lead to another
point within the triangle of parameters indicated in Fig.\
\ref{fig:merit15}. A better tolerance could be achieved if light
intensities are increased so that the gap between the two lines of
signal-to-noise and heating probability in Fig.\ \ref{fig:merit15}
broadens. However, in this case the potential depth would be
decreased, with increased loss of atoms through tunneling to the
disk surface.


\begin{figure}
\includegraphics[width=6.5cm]{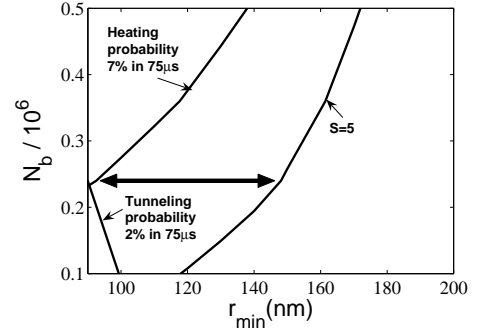}
\caption{Performance characteristics of a disk with diameter
$D=30$ $\mu$m and for a $75\mu$s integration time, at different
blue-detuned photon numbers and atom distances. The area confined
by the lines of $S=5$, heating probability $7\%$ and curve
represents tunneling probability of $2\%$ contains the parameter
region where stable atom trapping and detection can be achieved
with high confidence. The arrow represents the change in trap
center position if the intensity of the red- or blue-detuned light
is varied by $\pm 2\%$. \label{fig:merit30}}
\end{figure}

The same analysis with the contour lines for $S=5$ and heating
probability of 7\% for the disk diameter $D=30$ $\mu$m gives the
optimal point for atom detection within the parameter region of
Fig.\ \ref{fig:merit30} for $N_b=2.4 \times 10^5$ and a trap at
distance $r=120$ nm from the disk. In order to achieve atomic
trapping at this point we need $3.6\times 10^5$ red-detuned
photons in the disk. At this point we obtain $S=7$, the potential
depth is 1.6 mK and the harmonic oscillator frequency is
$(\omega_{\mathrm{x}}, \omega_{\mathrm{y}},\omega_{\mathrm{z}})=
2\pi\times (0.92,4.2,0.11)$ MHz. The ground state energy is
approximately 126 $\mu$K and the ground state dimensions are
$(\Delta x,\Delta y,\Delta z)=(15.9,7.4,46)$ nm. The heating
probability is 5.7\%. For these optimized parameters a change of
$\pm$2\% in either the blue or red-detuned light intensity will
still lead to another point within the area of parameters
indicated in Fig.\ \ref{fig:merit30}, i.e., the blue or
red-detuned intensities may be changed within $\pm$2\% while
maintaining $S>5$.


\section{Experimental feasibility and conclusions}
\label{sec:conclusions}

We have shown that a single atom can be trapped near the surface
of a microdisk resonator such that its presence can be detected
with negligible heating. The detection of the atom is done by a
blue-detuned WGM of the resonator, while trapping at a fixed
position is achieved by a second, red-detuned WGM. The two light
fields create a trapping potential at a distance of 100-150 nm
from the disk surface. At this distance, the atom-surface
attractive interaction (van-der-Waals force) is much weaker than
the light force, while the optical potential is sufficiently
strong to create a deep trap for the atom. The atom is then
confined in the radial direction and in the $z$ direction
(perpendicular to the chip surface). For trapping in the
tangential direction we suggest that the red light WGM is coupled
to the microdisk from both sides, such that a red-detuned standing
wave is formed along the disk perimeter and the atom may be
trapped in any of the maxima of the red-light.

The use of photonics for atom chips has been discussed recently
\cite{Birkl}. Detection of atoms by evanescent fields has been
achieved experimentally \cite{cornelussen}. Moreover, the use of
bichromatic light for guiding or trapping atoms has also been
discussed before
\cite{vern,ovchinnikov,mabuchi,barnett,burke,lekien,domrit}. The
idea of utilizing a two dimensional microsphere, i.e., a disk or
ring with a favorable fabrication feasibility, was put forward by
us in a recent paper \cite{rosenblit}. In this work, a realistic
tolerance analysis combining all the above ideas has been
presented. We show that with current fabrication capabilities, a
tunable high-Q device may be built to detect single atoms. More
specifically, we note that state-of-the-art fabrication has
reached a point, where previous concerns regarding the spatial
instability of the trapping potential due to light back scatter
from imperfections \cite{burke}, may now be no longer valid.
Concerning the required light mode stability we estimate an
acceptable tolerance of $\sim$2\%. This has been calculated when
demanding high stability for the trap parameters as well as
atom-light interaction, and is highly dependent on surface
roughness and mode coupling. Currently, surface roughness of order
1 nm is achieved by re-flow processes where the surface layer is
melted and allowed to reform itself with surface tension forces.
We have calculated that weak mode coupling and such roughness will
enable the above required light intensity accuracy.

\begin{figure}
\includegraphics[width=6.5cm]{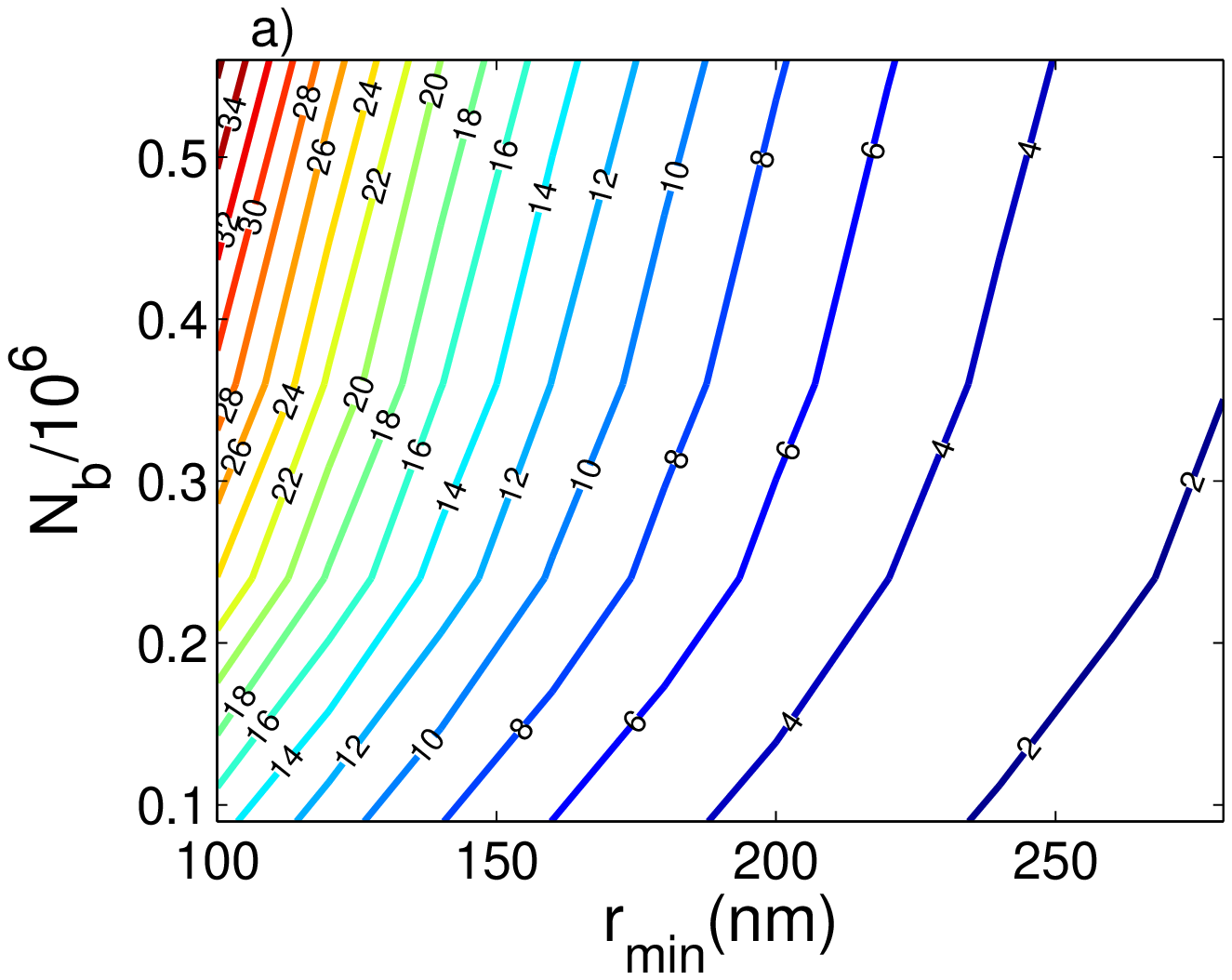}
\includegraphics[width=6.5cm]{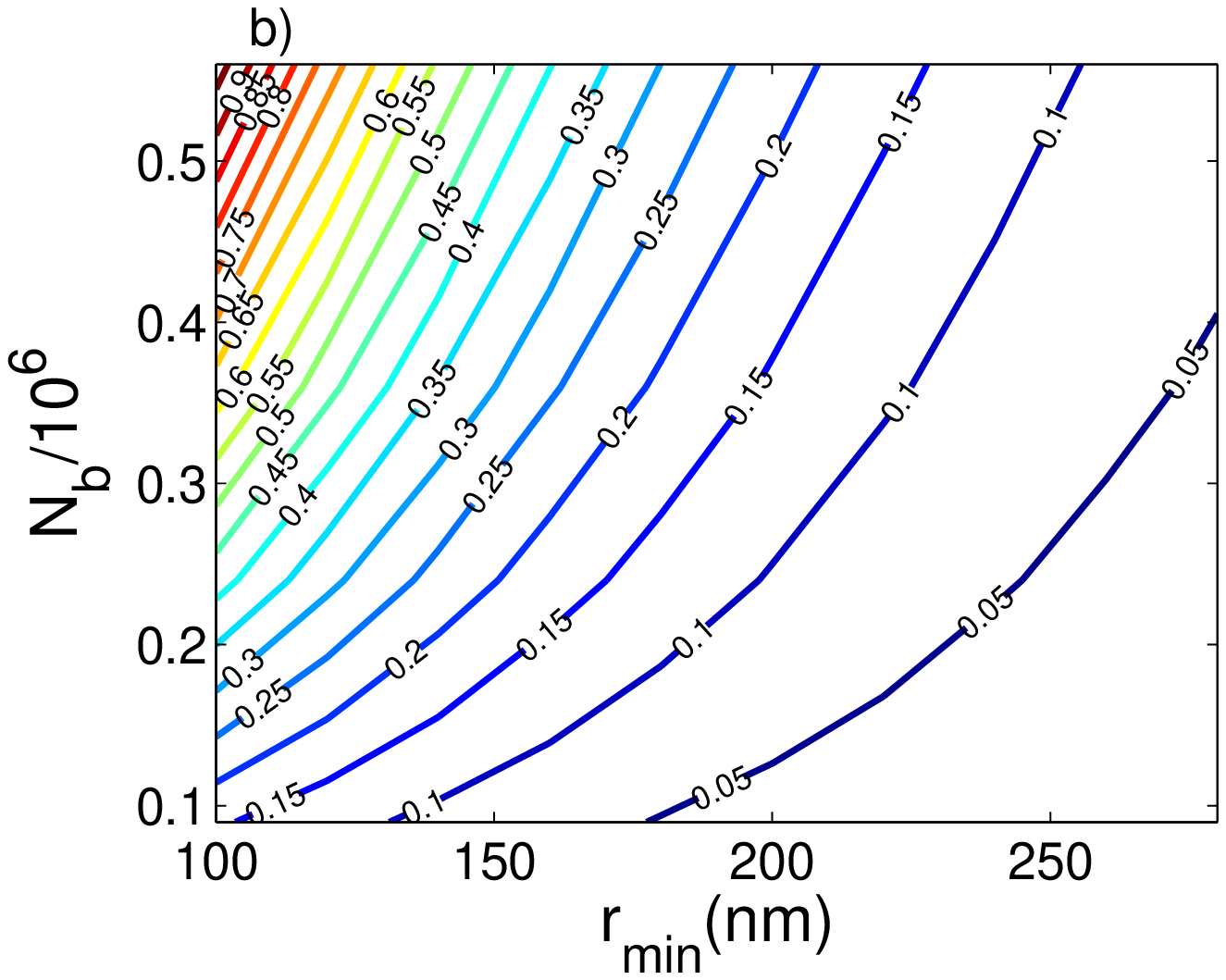}
  \caption{(a) Signal to noise ratio and (b) total scattered photon numbers
  vs. distance from the disk and blue detuned photon number for disk diameter 30 $\mu$m.
  Integration time is 5 $\mu$s, rms roughness $\sigma=0.2$ nm, and gap size $\sim$0.7 $\mu$m.}
  \label{fig:rms30}
\end{figure}

Recent experimental works give rise to the possibility of further
decreasing surface roughness to below 0.2 nm \cite{kklee,rabady}.
Applying such roughness to the microdisk we can decrease the total
scattering photon number to below one and consequently contemplate
the possibility of a true nondestructive measurement (Fig.\
\ref{fig:rms30}).

The intensity of the input field becomes important also with
regard to nonlinear effects. Decreasing the input light intensity
can weaken nonlinear mechanisms of instability such as Kerr
optical parametric oscillation \cite{kerr} and radiation pressure
induced mechanical oscillation \cite{vibration}.

Another possibility to achieve trapping during detection, is
through magnetic trapping. We have shown that magnetic trapping is
usually unsuitable for this as optical polarization control is
hard to achieve in WGMs and because light-induced Raman
transitions will transfer atoms into magnetically untrapped
states. Other possibilities such as by use of an attractive
electric field, may offer more stability, but have their own
drawbacks. For example, any metallic electrode near the atom would
produce thermally induced EM noise, not to mention absorb and
diffract the very sensitive mode of the high-Q resonator.
Nevertheless, such options are under investigation and will be
analyzed elsewhere.

Further work will also need to address in detail the issue of
loading, i.e, how the atoms are brought close to the disk surface.
The loading is crucial as it deals with the interplay between the
specific atom optics elements such as guides and traps, and the
detector responsible for extracting the signal. Here, evanescent
disk fields {\it above} the disk, may prove helpful.

Finally, this work described non-destructive detection in the
sense of negligible heating during the detection. Future work will
need to analyze another harmful mechanism in the form of Raman
transitions during the detection between the internal degrees of
freedom, i.e., between hyperfine states. It is crucial for quantum
technology that detection will not alter the hyperfine state
occupancy, as the latter form in most cases the observable of the
quantum optics operations.

Let us conclude by stating that we have shown that indeed the road
is open for the fabrication of an atom chip in which a micro disk
resonator would be integrated. Such an apparatus may offer insight
into new experimental regimes, while in parallel devices for
quantum technology may be realized.





\begin{thebibliography}{99}
\bibitem{equation} The intensity change due to the additional counterpropagating
mode is given by $I_+ \rightarrow$ $I_+[1 + I_-/I_+ +
2(I_/I_+)^{1/2}cos(\varphi)]$, where $\varphi$ is an arbitrary
relative phase.  If the counterpropagating mode intensity is
0.01\% of the propagating mode then the sum intensity is changed
by 2\%.
\end{thebibliography}
\end{document}